\documentstyle[12pt,prd,aps,epsfig,preprint]{revtex}
\def\beq{\begin{equation}}
\def\eeq{\end{equation}}
\def\bea{\begin{eqnarray}}
\def\eea{\end{eqnarray}}
\def\1{\c{c}}
\def\2{\c{C}}
\def\3{\u{g}}
\def\4{\u{G}}
\def\5{{\i}}
\def\6{\.{I}}
\def\7{\"{o}}
\def\8{\"{O}}
\def\9{\c{s}}
\def\0{\c{S}}
\def\*{\"{u}}
\def\,{\"{U}}
\begin{document}
\draft
\date{\today}
\title{Different Contributions in $\omega\rightarrow \pi^0\eta\gamma$ and
$\rho\rightarrow \pi^0\eta\gamma$ decays}

\author{Ayse Kucukarslan~\thanks{akucukarslan@metu.edu.tr}}
\address{ {\it Middle East Technical University, Physics Department,
06531 Ankara, Turkey}}
\author{Saime Solmaz~\thanks{skerman@balikesir.edu.tr}}
\address{ {\it Balikesir University, Physics Department, 10100,
Balikesir, Turkey}}

\maketitle
\begin{abstract}
We examine the radiative $\omega\rightarrow\pi^{0}\eta\gamma$ and
$\rho\rightarrow\pi^{0}\eta\gamma$ decays in a phenomenological
framework. We consider the VMD mechanism, chiral loops,
intermediate $a_{0}$-meson and $\rho-\omega$ mixing. We find the
values of the decay width coming from the different amplitudes and
compare the results with other studies. We observe that
$a_{0}$-meson intermediate state is very important in the case of
the $\rho\rightarrow\pi^{0}\eta\gamma$ decay and small in the
other case for which VMD contribution is dominant.
\end{abstract}

\thispagestyle{empty} ~~~~\\ 
\newpage
\setcounter{page}{1}

\section{INTRODUCTION}
Radiative decays of low-mass vector mesons into a single photon
and a pair of neutral pseudoscalars have attracted continuous
attention. The studies of such decays have been a case for tests
of vector meson dominance (VMD), through the sequential mechanism
$V\rightarrow PV\rightarrow PP\gamma$ \cite{R1,R2}. They also
offer the possibility of obtaining information on the nature of
low-mass scalar mesons. In particular the nature and the quark
substructure of the two scalar mesons, isoscalar $f_{0} (980)$ and
isovector $a_{0} (980)$, have not been established yet. Several
proposals have been made about the nature of these states:
$q\bar{q}$ states in quark model \cite{R3}, $K\bar{K}$ molecules
\cite{R4} or multiquark $q^{2}\bar{q^2}$ states \cite{R5,R6}.

Theoretical study of $\omega$ and $\rho$ meson decays into a
single photon and pseudoscalar $\pi^{0}$ and $\eta$ mesons as well
as other radiative vector meson decays was initiated by Fajfer and
Oakes \cite{R7}. They described these decays by the gauged
Wess-Zumino terms in a low-energy effective Lagrangian and
calculated the branching ratios for these decays in which scalar
meson contributions were neglected. In their study, they obtained
the following branching ratios for
$\omega\rightarrow\pi^{0}\eta\gamma$ and
$\rho\rightarrow\pi^{0}\eta\gamma$ decays:
$BR(\omega\rightarrow\pi^{0}\eta\gamma)=6.26\times10^{-6}$,
$BR(\rho\rightarrow\pi^{0}\eta\gamma)=3.98\times10^{-6}$. The
contributions of intermediate vector mesons to the decays
$V^{0}\rightarrow P^{0}P^{0}\gamma$ were later considered by
Bramon et al. \cite{R2} using standart Lagrangians obeying SU(3)
symmetry. Their results for the decay rates and the branching
ratios of the decays $\omega\rightarrow\pi^{0}\eta\gamma$ and
$\rho\rightarrow\pi^{0}\eta\gamma$ were
$\Gamma(\omega\rightarrow\pi^{0}\eta\gamma)=1.39~eV$,
$BR(\omega\rightarrow\pi^{0}\eta\gamma)=1.6\times10^{-7}$ and
$\Gamma(\rho\rightarrow\pi^{0}\eta\gamma)=0.061~eV$,
$BR(\rho\rightarrow\pi^{0}\eta\gamma)=4\times10^{-10}$. Their
results were not incompatible with those by Fajfer and Oakes
\cite{R7} even if the initial expressions for the Lagrangians were
the same. Later, Bramon et al. \cite{R8} studied these decays
within the framework of chiral effective Lagrangians enlarged to
include on-shell vector mesons using chiral perturbation theory,
and they calculated the branching ratios for
$\omega\rightarrow\pi^{0}\eta\gamma$ and
$\rho\rightarrow\pi^{0}\eta\gamma$ decays as well as other
radiative vector meson decays of the type $V^{0}\rightarrow
P^{0}P^{0}\gamma$ at the one loop level. They showed that the one
loop contributions are finite and to this order no counterterms
are required. In this approach, the decays
$\omega\rightarrow\pi^{0}\eta\gamma$ and
$\rho\rightarrow\pi^{0}\eta\gamma$ proceed through the
intermediate vector meson states and the charged kaon loops and
they obtained the contributions of charged kaon-loops to the decay
rates of these decays as
$\Gamma(\omega\rightarrow\pi^{0}\eta\gamma)_{K}=0.013~eV$ and
$\Gamma(\rho\rightarrow\pi^{0}\eta\gamma)_{K}=0.006~eV$ with the
pion-loop contributions vanishing in the good isospin limit. Their
analysis showed that kaon-loop contributions are one or two orders
of magnitude smaller than the VMD contributions and the dominant
pion-loops are forbidden in these decays due to isospin symmetry.
These decays were also investigated by Prades \cite{R9}. Using
chiral Lagrangians and the extended Nambu-Jona-Lasinio model he
calculated the branching ratios for these decays. The branching
ratios for the radiative $\omega\rightarrow\pi^{0}\eta\gamma$ and
$\rho\rightarrow\pi^{0}\eta\gamma$ decays were found as
$BR(\omega\rightarrow\pi^{0}\eta\gamma)=8.3\times10^{-8}$ and
$BR(\rho\rightarrow\pi^{0}\eta\gamma)=2.0\times10^{-10}$.
Furthermore, the radiative $\omega\rightarrow\pi^{0}\eta\gamma$
and $\rho\rightarrow\pi^{0}\eta\gamma$ decays were also considered
by Gokalp et al. \cite{R10} taking into account the contributions
of intermediate $a_{0}$-meson and intermediate vector meson
states. The decay rates and the branching ratios for
$\omega\rightarrow\pi^{0}\eta\gamma$ and
$\rho\rightarrow\pi^{0}\eta\gamma$ decays, they obtained, were
$\Gamma(\omega\rightarrow\pi^{0}\eta\gamma)=1.62~eV$,
$BR(\omega\rightarrow\pi^{0}\eta\gamma)=1.92\times10^{-7}$ and
$\Gamma(\rho\rightarrow\pi^{0}\eta\gamma)=0.43~eV$,
$BR(\rho\rightarrow\pi^{0}\eta\gamma)=2.9\times10^{-9}$. They
concluded that although $a_{0}$-meson intermediate state amplitude
makes a small contribution to $\omega\rightarrow\pi^{0}\eta\gamma$
decay it makes a substantial contribution to
$\rho\rightarrow\pi^{0}\eta\gamma$ decay. Recently, the radiative
decays of the $\rho$ and $\omega$ mesons into two neutral mesons,
$\pi^{0}\pi^{0}$ and $\pi^{0}\eta$, including the mechanisims of
sequential vector meson decay, $\rho-\omega$ mixing and chiral
loops have been studied by Palomar et al. \cite{R11}. They
obtained the branching ratios for the decays
$\omega\rightarrow\pi^{0}\eta\gamma$ and
$\rho\rightarrow\pi^{0}\eta\gamma$ as
$BR(\omega\rightarrow\pi^{0}\eta\gamma)=3.3\times10^{-7}$,
$BR(\rho\rightarrow\pi^{0}\eta\gamma)=7.5\times10^{-10}$ and noted
that, the dominant contribution is the one corresponding to the
sequential mechanism for both cases. Indeed, in their study the
$\rho-\omega$ mixing was found non negligible for
$\omega\rightarrow\pi^{0}\eta\gamma$ and
$\rho\rightarrow\pi^{0}\eta\gamma$ decays.

Theoretically, the effects of the $\rho-\omega$ mixing in the
$\omega\rightarrow\pi^{0}\eta\gamma$ and
$\rho\rightarrow\pi^{0}\eta\gamma$ decays have not been studied
extensively up to now. One of the rare studies of these decays was
by Palomar et al. \cite{R11}. In their study the chiral loops were
obtained using elements of $U\chi PT$ which lead to the excitation
of the scalar resonances without the need to include them
explicitly in the formalism. However, in our work the effect of
$a_{0}$ (980) meson in the decay mechanisms of
$\omega\rightarrow\pi^{0}\eta\gamma$ and
$\rho\rightarrow\pi^{0}\eta\gamma$ decays is included as resulting
from $a_{0}$-pole intermediate state. We study these decays within
the framework of a phenomenological approach in which the
contributions of intermediate vector meson states, chiral loops,
$\rho-\omega$-mixing and of scalar $a_{0}$ (980) intermediate
meson state are considered. Expressions related with branching
ratios for $\omega\rightarrow\pi^{0}\eta\gamma$ and
$\rho\rightarrow\pi^{0}\eta\gamma$ decays are presented in
conclusion.

\section{VMD Contributions}

In our calculation we use the Feynman diagrams, corresponding to
this mechanism, shown in Fig. 1a for $\omega\rightarrow
\pi^0\eta\gamma$ decay and in Fig. 2a for $\rho\rightarrow
\pi^0\eta\gamma$ decay. The Lagrangian for the
$\omega\rho\pi$-vertex takes the following form
\begin{eqnarray} \label{e1}
{\cal L}^{eff}_{\omega\rho\pi}=g_{\omega\rho\pi}
\epsilon^{\mu\nu\alpha\beta}\partial_{\mu}\omega_{\nu}
\partial_{\alpha}\vec{\rho}_{\beta}\cdot\vec{\pi}~~.
\end{eqnarray}
Since the coupling constant $g_{\omega\rho\pi}$ can not be
determined directly from experiments, theoretically it is
extracted from some models and obtained the values between $11$
$GeV^{-1}$ and $16$ $GeV^{-1}$. We use the value as $15$
$GeV^{-1}$ for this coupling constant in this work. The
$V\varphi\gamma$-vertices come from the Lagrangians
\begin{eqnarray}\label{e2}
{\cal L}^{eff}_{V\varphi\gamma}=g_{V\varphi\gamma}
\epsilon^{\mu\nu\alpha\beta}\partial_{\mu}V_{\nu}
\partial_{\alpha}A_{\beta}\varphi~~,
\end{eqnarray}
where $V_{\nu}$ is the vector meson field $\omega_{\nu}$ or
$\rho_{\nu}$, $\varphi$ is the pseudoscalar field $\pi^{0}$ or
$\eta$, and $A_{\beta}$ is the photon field. Using the
experimental partial widths for
$\Gamma(V\rightarrow\pi^{0}\gamma)$ and
$\Gamma(V\rightarrow\eta\gamma)$ \cite{R12}, we determine the
coupling constants as $g_{\rho\pi\gamma}$=0.696,
$g_{\rho\eta\gamma}$=1.171, $g_{\omega\pi\gamma}$=1.821, and
$g_{\omega\eta\gamma}$=0.400. For the $VV\eta$-vertex we use the
following effective Lagrangian
\begin{eqnarray}\label{e3}
{\cal L}^{eff}_{VV\eta}=g_{VV\eta} \epsilon^{\mu\nu\alpha\beta}
\partial_{\mu}V_{\nu}V_{\alpha}\partial_{\beta}\eta~~.
\end{eqnarray}
Utilizing the experimental decay widths of the
$\omega\rightarrow3\pi$ and $\phi\rightarrow3\pi$ decays, Klingl
et al. \cite{R13} obtained the coupling constant $g_{VV\eta}$ as
$g_{\omega\omega\eta}=g_{\rho\rho\eta}=2.624$ $GeV^{-1}$.

We also use the following momentum dependent width, as discussed
by O'Connell et al. \cite{R14} for $V=\rho$ or $\omega$ meson
\begin{eqnarray} \label{e4}
\Gamma_{V}(q^{2})=\Gamma_{V}\frac{M_{V}}{\sqrt{q^{2}}}\left(\frac{q^{2}-4M_{\pi}^{2}}{M_{V}^{2}-4M_{\pi}^{2}}\right)^{3/2}
~\theta(q^{2}-4M_{\pi}^{2})~~.
\end{eqnarray}
Since $\Gamma_{\omega}$ is small, this effect is negligible for
$\rho\rightarrow \pi^0\eta\gamma$ decay.

\section{Chiral loop contributions}

Apart from the VMD contributions, there is another mechanism based
on the chiral kaon-loop whose contribution is quite small in the
two cases, $\omega\rightarrow\pi^{0}\eta\gamma$ and
$\rho\rightarrow\pi^{0}\eta\gamma$. In spite of this, we add the
kaon-loop contribution for completeness in our calculation. This
mechanism has been studied in \cite{R8,R9,R11} for these decays
and here we follow closely results of these studies.

The one loop Feynman diagrams for
$\omega\rightarrow\pi^{0}\eta\gamma$ and
$\rho\rightarrow\pi^{0}\eta\gamma$ are of the form shown in Fig.
1b and Fig. 2b, respectively. For the contribution of these
diagrams we use the amplitude given in Ref. \cite{R8} derived
using chiral perturbation theory. The amplitude is
\begin{eqnarray} \label{e8}
{\cal A}(V\rightarrow\pi^{0}\eta\gamma)_{K}=
-\frac{eg}{6\sqrt{3}\pi^{2}f_{\pi}^{2}}(3p^{2}-6k\cdot
p-4M_{K}^{2}) [(\epsilon\cdot u)(k\cdot p)-(\epsilon\cdot p)
(k\cdot u)]\frac{1}{M_{K}^{2}}I(a,b)
\end{eqnarray}
where $I(a,b)$ is the loop function defined as
\begin{eqnarray} \label{e9}
I(a,b)=\frac{1}{2(a-b)}-\frac{2}{(a-b)^{2}}\left[f\left(\frac{1}{b}\right)-
f\left(\frac{1}{a}\right)\right]+\frac{a}{(a-b)^{2}}
\left[g\left(\frac{1}{b}\right)-g\left(\frac{1}{a}\right)\right]
\end{eqnarray}
where $a=M_{V}^{2}/M_{K}^{2}$, $b=(p-k)^{2}/M_{K}^{2}$, $g\simeq
4.2$ and $f_{\pi}=132$ MeV, f(x) and g(x) are defined in
\cite{R15} which were evaluated by Lucio and Pestieau.

We will not consider the pion loops for these decays because it
does not contribute in good isospin limit.

\section{Scalar meson Contributions}

We add $a_{0}$-meson as an intermediate state to the decay
mechanism of these decays. The scalar $a_{0}$-meson contribution
were studied before \cite{R10,R16} by Gokalp et al. within the
framework of a phenomenological approach for vector meson decays.

We use the Feynman diagrams for
$\omega\rightarrow\pi^{0}\eta\gamma$ and
$\rho\rightarrow\pi^{0}\eta\gamma$ decays as shown in Fig. 1c and
Fig. 2c, respectively. The vertices, $V a_{0}\gamma$ and
$a_{0}\pi^{0}\eta$, come from the Lagrangians
\begin{eqnarray} \label{e10}
{\cal
L}^{eff}_{Va_{0}\gamma}=g_{Va_{0}\gamma}(\partial^{\alpha}V^{\beta}\partial_{\alpha}A_{\beta}-
\partial^{\alpha}V^{\beta}\partial_{\beta}A_{\alpha})a_{0}
\end{eqnarray}
\begin{eqnarray} \label{e11}
{\cal
L}^{eff}_{a_{0}\pi\eta}=g_{a_{0}\pi\eta}\vec{\pi}\cdot\vec{a_{0}}\eta~~,
\end{eqnarray}
where we also define the coupling constants $g_{Va_{0}\gamma}$ and
$g_{a_{0}\pi\eta}$. Since there are no direct experimental results
for $Va_{0}\gamma$-vertex, we use the values for the coupling
constants $g_{Va_{0}\gamma}$ as $g_{\rho
a_{0}\gamma}=(1.69\pm0.39)$ $GeV^{-1}$ and $g_{\omega
a_{0}\gamma}=(0.58\pm0.13)$ $GeV^{-1}$ which was determined using
the QCD sum rule method in \cite{R17}. The decay rate for the
$a_{0}\rightarrow\pi^{0}\eta$ decay resulting from the above
Lagrangian is
\begin{eqnarray} \label{e12}
\Gamma(a_{0}\rightarrow\pi^{0}\eta)=\frac{g_{a_{0}\pi\eta}^{2}}{16\pi
M_{a_{0}}}\sqrt{\left[1-\frac{(M_{\pi^{0}}+M_{\eta})^{2}}{M_{a_{0}}^{2}}\right]
\left[1-\frac{(M_{\pi^{0}}-M_{\eta})^{2}}{M_{a_{0}}^{2}}\right]}~~.
\end{eqnarray}
Using the value $\Gamma_{a_{0}}=(0.069\pm0.011)$ GeV \cite{R18},
we obtain the coupling constant $g_{a_{0}\pi\eta}$ as
$g_{a_{0}\pi\eta}=(2.32\pm0.18)$ GeV. We use energy-dependent
width for the intermediate $a_{0}$-meson in the propagators, which
leads to an increase of the decay width when compared to the
calculation done with a constant width. The energy-dependent width
for $a_{0}$-meson is
\begin{eqnarray} \label{e13}
\Gamma_{a_{0}}(q^{2})=\Gamma_{a_{0}}\frac{M_{a_{0}}^{3}}{(q^{2})^{3/2}}
\sqrt{\frac{\left[q^{2}-(M_{\pi^{0}}+M_{\eta})^{2}\right]\left[q^{2}-(M_{\pi^{0}}-M_{\eta})^{2}\right]}
{\left[M_{a_{0}}^{2}-(M_{\pi^{0}}+M_{\eta})^{2}\right]\left[M_{a_{0}}^{2}-(M_{\pi^{0}}-M_{\eta})^{2}\right]}}
~~.
\end{eqnarray}

\section{The Effects of $\rho-\omega$ Mixing}

In addition to the VMD contribution given in section 2, we also
consider the mixing of the $\rho$ and $\omega$ mesons which is
constituted isospin violation effect due to mass differences of
quark and the electromagnetic interaction. This mixing has been
extracted from an analysis of
$e^{+}e^{-}\rightarrow\pi^{+}\pi^{-}$ in the $\rho-\omega$
interference region. Guetta and Singer \cite{R19} firstly
considered the $\rho-\omega$ mixing in the vector meson decays and
then it was used by Bramon et al. \cite{R20} and Palomar et al.
\cite{R11}. New contribution coming from the $\rho-\omega$ mixing
is to add to the intermediate vector meson diagrams of Fig. 1a and
Fig. 2a for $\omega\rightarrow\pi^{0}\eta\gamma$ and
$\rho\rightarrow\pi^{0}\eta\gamma$, respectively, expressing the
mixing between the isospin states which is described by adding to
the effective Lagrangian a term ${\cal
L}=\prod_{\rho\omega}\omega_{\mu}\rho_{\mu}$ leading to the
physical states $\rho=\rho(I=1)+\varepsilon\omega(I=0)$ and
$\omega=\omega(I=0)-\varepsilon\rho(I=1)$. Then the full amplitude
to be written as
\begin{eqnarray}\label{e14}
\label{e5} {\cal A}(V \rightarrow\pi^{0}\eta\gamma)= {\cal
A}_{0}(V \rightarrow\pi^{0}\eta\gamma)+\varepsilon\widetilde{\cal
A}(V' \rightarrow\pi^{0}\eta\gamma)~~,
\end{eqnarray}
where ${\cal A}_{0}$ and $\widetilde{\cal A}$ include the
contributions coming from the different terms, $\varepsilon$ is
the mixing parameter
\begin{eqnarray}\label{e15}
\varepsilon\equiv\frac{\prod_{\rho\omega}}{M_{V}^{2}-M_{V'}^{2}-i(M_{V}\Gamma_{V}-M_{V'}\Gamma_{V'})}
\end{eqnarray}
and it is obtained as $\varepsilon$=(-0.006+i0.036) using the
experimental values for $M_{V}$ and $\Gamma_{V}$ and
$\prod_{\rho\omega}$=$(-3811\pm370)$ $MeV^{2}$ which determined by
O'Connell et al. \cite{R14}.

Another effect of the mixing is modifying the propagator in ${\cal
A}_{0}$ as follow
\begin{eqnarray} \label{e16}
\frac{1}{D_{V}(s)}\longrightarrow\frac{1}{D_{V}(s)}\left(1+
\frac{g_{V'\pi\gamma}}{g_{V\pi\gamma}}\frac{\prod_{\rho\omega}}{D_{V'}(s)}\right)
\end{eqnarray}
with $D_{V}(s)=s-M_{V}^{2}+iM_{V}\Gamma_{V}$.

We express the invariant amplitude ${\cal A}(E_{\gamma},E_{\pi})$
using the $\rho-\omega$-mixing for the
$\omega\rightarrow\pi^{0}\eta\gamma$ as ${\cal
A}(\omega\rightarrow\pi^{0}\eta\gamma)={\cal
A}^{0}(\omega\rightarrow\pi^{0}\eta\gamma)+\varepsilon\widetilde{\cal
A}(\rho\rightarrow\pi^{0}\eta\gamma)$ where ${\cal A}^{0}$ and
$\widetilde{\cal A}$ are the invariant amplitudes coming from the
diagrams (a), (b), (c) in Fig. 1 and in Fig. 2. For the
$\rho\rightarrow\pi^{0}\eta\gamma$ decay, we follow the same
procedure and we can write the full amplitude as ${\cal
A}(\rho\rightarrow\pi^{0}\eta\gamma)={\cal
A}^{0}(\rho\rightarrow\pi^{0}\eta\gamma)-\varepsilon\widetilde{\cal
A}(\omega\rightarrow\pi^{0}\eta\gamma)$.

In our calculation, the decay width for these decays can be
obtained by integration
\begin{eqnarray} \label{e17}
\Gamma(V\rightarrow\pi^{0}\eta\gamma)=\int_{E_{\gamma,min.}}^{E_{\gamma,max.}}dE_{\gamma}
\int_{E_{\pi,min.}}^{E_{\pi,max.}}dE_{\pi}\frac{d\Gamma}{dE_{\gamma}dE_{\pi}}~~.
\end{eqnarray}
The minimum photon energy is $E_{\gamma , min}$=0 and the maximum
photon energy is given as $E_{\gamma
,max.}=[M_{V}^{2}-(M_{\pi}+M_{\eta})^{2}]/2M_{V}$. The minimum and
maximum values for pion energy $E_{\pi}$ are given by
\begin{eqnarray}\label{e18}
&&\frac{1}{2(2E_\gamma M_{V}-M_{V}^{2})}~\{ -2E_\gamma^2
M_{V}-M_{V}(M_{V}^2+M_{\pi}^2-M_\eta^2)+E_\gamma(3M_{{V}}^2+M_{\pi}^2-M_\eta^2)
\nonumber
\\&& \pm E_\gamma
[~4E_\gamma^2
M_{V}^2+M_{V}^4+(M_{\pi}^2-M_\eta^2)^2-2M_{V}^2(M_{\pi}^2+M_\eta^2)
\nonumber \\ ~~&&+4E_\gamma
M_{V}(-M_{V}^2+M_{\pi}^2+M_\eta^2)~]^{1/2}~\}~~.
\end{eqnarray}
The differential decay probability of
$V^{0}\rightarrow\pi^{0}\eta\gamma$ decay for an unpolarized
$V^{0}$-meson $(V^{0}=\omega,\rho^{0})$ at rest is then given as
in terms of the invariant amplitude $A(E_{\gamma}, E_{\pi})$
\begin{eqnarray} \label{e19}
\frac{d\Gamma}{dE_{\gamma}dE_{\pi}}=\frac{1}{(2\pi)^{3}}\frac{1}{8M_{V}}|{\cal
A}|^{2}
\end{eqnarray}
where $E_{\gamma}$ and $E_{\pi}$ are the photon and pion energies
respectively. We perform an average over the spin states of the
vector meson and a sum over the polarization states of the photon.

\section{Results and Conclusion}

The contributions of different amplitudes to the decay rate and
the branching ratio of the decays,
$\omega\rightarrow\pi^{0}\eta\gamma$,
$\rho\rightarrow\pi^{0}\eta\gamma$, are shown in Table 1 and Table
2, respectively. We consider the intermediate vector meson, chiral
loops, intermediate $a_{0}$-meson and $\rho-\omega$ mixing. The
dominant contribution is the one corresponding to the vector meson
dominance mechanism in two cases except for the
$\rho\rightarrow\pi^{0}\eta\gamma$ decay. On the contrary,
intermediate $a_{0}$-meson is the dominant contribution of the
$\rho\rightarrow\pi^{0}\eta\gamma$ decay.

The resulting photon spectra for the decay rate is plotted in Fig.
3 for the decay $\omega\rightarrow\pi^{0}\eta\gamma$ and in Fig. 4
for the decay $\rho\rightarrow\pi^{0}\eta\gamma$. The separate
contributions coming from vector meson dominance amplitude,
$\omega-\rho$-mixing amplitude, $a_{0}$-meson intermediate state
amplitude, chiral loop amplitudes and their interference, as well
as the contribution of total amplitude are explicitly shown. As we
can see in two figures, the contribution of the VMD amplitude does
not change if we add the effect of $\omega-\rho$-mixing. The
situation changes in two cases when we include VMD, $a_{0}$-meson
intermediate state amplitude with $\omega-\rho$-mixing. The
interference term between all contribution is constructive for
$\omega\rightarrow\pi^{0}\eta\gamma$ decay as shown in Fig. 3. For
the decay $\rho\rightarrow\pi^{0}\eta\gamma$, $a_{0}$-meson
intermediate state amplitude contribution is quite significant in
comparison with other contributions as seen clearly in Fig. 4.

The effects of the $\rho-\omega$ mixing are too small for both
cases, especially for $\rho\rightarrow\pi^{0}\eta\gamma$ decay
it's not any contribution, but it modifies the propagator in the
vector meson dominance mechanism, so the $\rho-\omega$ mixing
should be added the calculation. The loop contribution for the
$\omega\rightarrow\pi^{0}\eta\gamma$ and
$\rho\rightarrow\pi^{0}\eta\gamma$ decays is found small due to
the relatively high mass of the kaons as mentioned also in
\cite{R8}.

In Table 3, we collect the results of other analysis and compare
our results with other studies and experiment. In Ref. \cite{R7}
an approach with low energy effective Lagrangians with gauged
Wess-Zumino terms was followed. A different procedure was followed
in \cite{R9} using chiral Lagrangians and the extended
Nambu-Jona-Lasinio model to fix the couplings of the resonance
contribution. The vector meson dominance mechanisms were
considered only in \cite{R2} and then the results were improved in
\cite{R8} for the $V^{0}\rightarrow P^{0}P^{0}\gamma$ decays using
the chiral perturbation theory. In \cite{R11} VMD, chiral loops
obtained using elements of unitarized chiral perturbation theory
applied in the study of meson-meson interaction, and $\rho-\omega$
mixing were considered for $V\rightarrow PP\gamma$ decays. Then,
in \cite{R10} only VMD and intermediate $a_{0}$-meson were
considered within the framework of a phenomenological approach.

Since we don't have any experimental value the results obtained
for the branching ratio of the $\rho\rightarrow\pi^{0}\eta\gamma$
decay can not be compared with measurements. It should be expected
that in the near future experiments related with
$\rho\rightarrow\pi^{0}\eta\gamma$ decay will verify or refute our
results. For the case of the $\omega\rightarrow\pi^{0}\eta\gamma$
decay, recently the CMD-2 collaboration obtained the following
upper limit, $BR(\omega\rightarrow\pi^{0}\eta\gamma)<3.3\times
10^{-5}$ \cite{R21}. Therefore, evaluated values are in agreement
with the experimental limit for
$\omega\rightarrow\pi^{0}\eta\gamma$ decay.

\section{Acknowledgments}

We are grateful to A. Gokalp and O. Yilmaz for useful discussions.

\begin{table}
\caption{The decay widths coming from the different contributions
to the $\omega\rightarrow\pi^{0}\eta\gamma$ and
$\rho\rightarrow\pi^{0}\eta\gamma$.}
\begin{tabular}{|c|c|c|c|c|c|}
$\Gamma$ (eV) & VMD & VMD+$(\rho-\omega)$ mixing & K-loop &
$a_{0}-meson$ & Total \\ \hline
$\omega\rightarrow\pi^{0}\eta\gamma$ & 1.51 & 1.54
& 0.0133 & 0.70 & 4.83\\
\hline $\rho\rightarrow\pi^{0}\eta\gamma$ & 0.08 & 0.08 & 0.006 &
3.43 & 3.44\\
\end{tabular}
\end{table}

\begin{table}
\caption{The branching ratios coming from the different
contributions to the $\omega\rightarrow\pi^{0}\eta\gamma$ and
$\rho\rightarrow\pi^{0}\eta\gamma$.}
\begin{tabular}{|c|c|c|c|c|c|}
BR & VMD & VMD+$(\rho-\omega)$ mixing & K-loop & $a_{0}-meson$ & Total \\
\hline $\omega\rightarrow\pi^{0}\eta\gamma$ & $1.79\times10^{-7}$
& $1.82\times10^{-7}$
& $1.6\times10^{-9}$ & $8.25\times10^{-8}$ & $5.72\times10^{-7}$\\
\hline $\rho\rightarrow\pi^{0}\eta\gamma$ & $5.27\times10^{-10}$ &
$5.27\times10^{-10}$ & $4.0\times10^{-11}$ &
$2.3\times10^{-8}$ & $2.3\times10^{-8}$\\
\end{tabular}
\end{table}

\begin{table}
\caption{The branching ratios of the
$\omega\rightarrow\pi^{0}\eta\gamma$ and
$\rho\rightarrow\pi^{0}\eta\gamma$ decays in the literature.}
\begin{tabular}{|c|c|c|}
WORK & $\omega\rightarrow\pi^{0}\eta\gamma$ &
$\rho\rightarrow\pi^{0}\eta\gamma$ \\
\hline \cite{R7} & $6.26\times10^{-6}$ & $3.98\times10^{-6}$\\
\hline \cite{R9} & $8.3\times10^{-8}$ & $2.0\times10^{-10}$\\
\hline \cite{R2} & $1.6\times10^{-7}$ & $4.0\times10^{-10}$\\
\hline \cite{R8} & $1.6\times10^{-7}$ & $4.0\times10^{-10}$\\
\hline \cite{R11} & $3.3\times10^{-7}$ & $7.5\times10^{-10}$\\
\hline \cite{R10} & $1.92\times10^{-7}$ & $2.9\times10^{-9}$\\
\hline this work & $5.72\times10^{-7}$ & $2.3\times10^{-8}$\\
\hline experiment & $< 3.3\times10^{-5}$ & --\\
\end{tabular}
\end{table}

\newpage
\begin{figure}\vspace*{0.0cm}\hspace{1.5cm}
\epsfig{figure=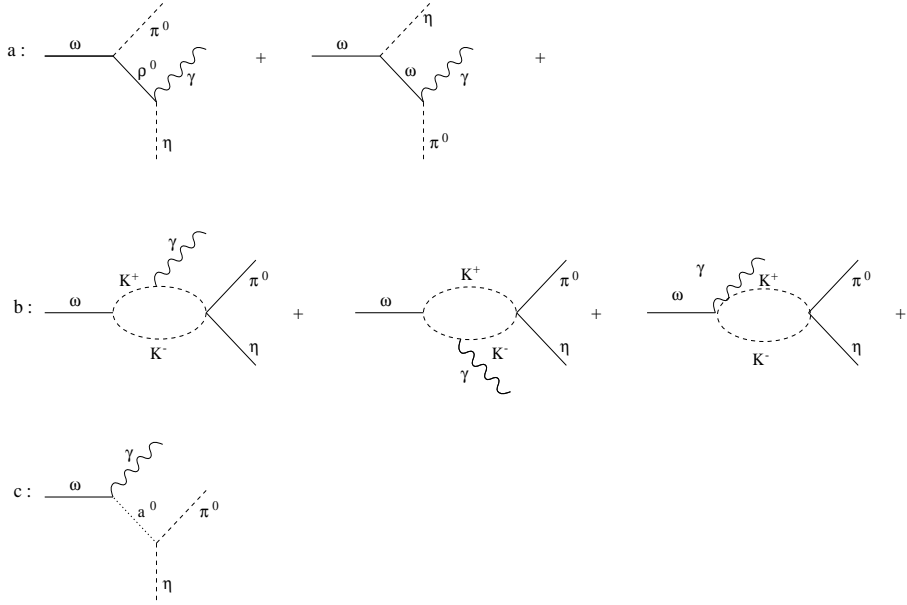,width=8cm,height=12cm,
angle=270}\vspace*{0.5cm} \caption{Feynman diagrams for the decay
$\omega\rightarrow\pi^0\eta\gamma$.} \label{fig1}
\end{figure}

\begin{figure}
\vspace*{0.0cm}\hspace{1.5cm}
\epsfig{figure=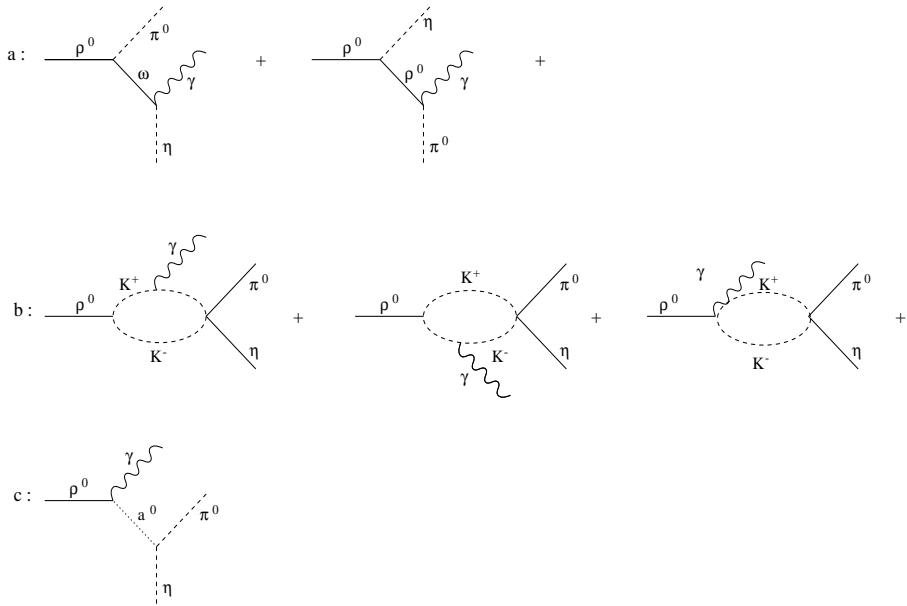,width=8cm,height=12cm,
angle=270}\vspace*{0.5cm} \caption{Feynman diagrams for the decay
$\rho^0\rightarrow\pi^0\eta\gamma$.} \label{fig2}
\end{figure}

\newpage
\begin{figure}
\epsfig{figure=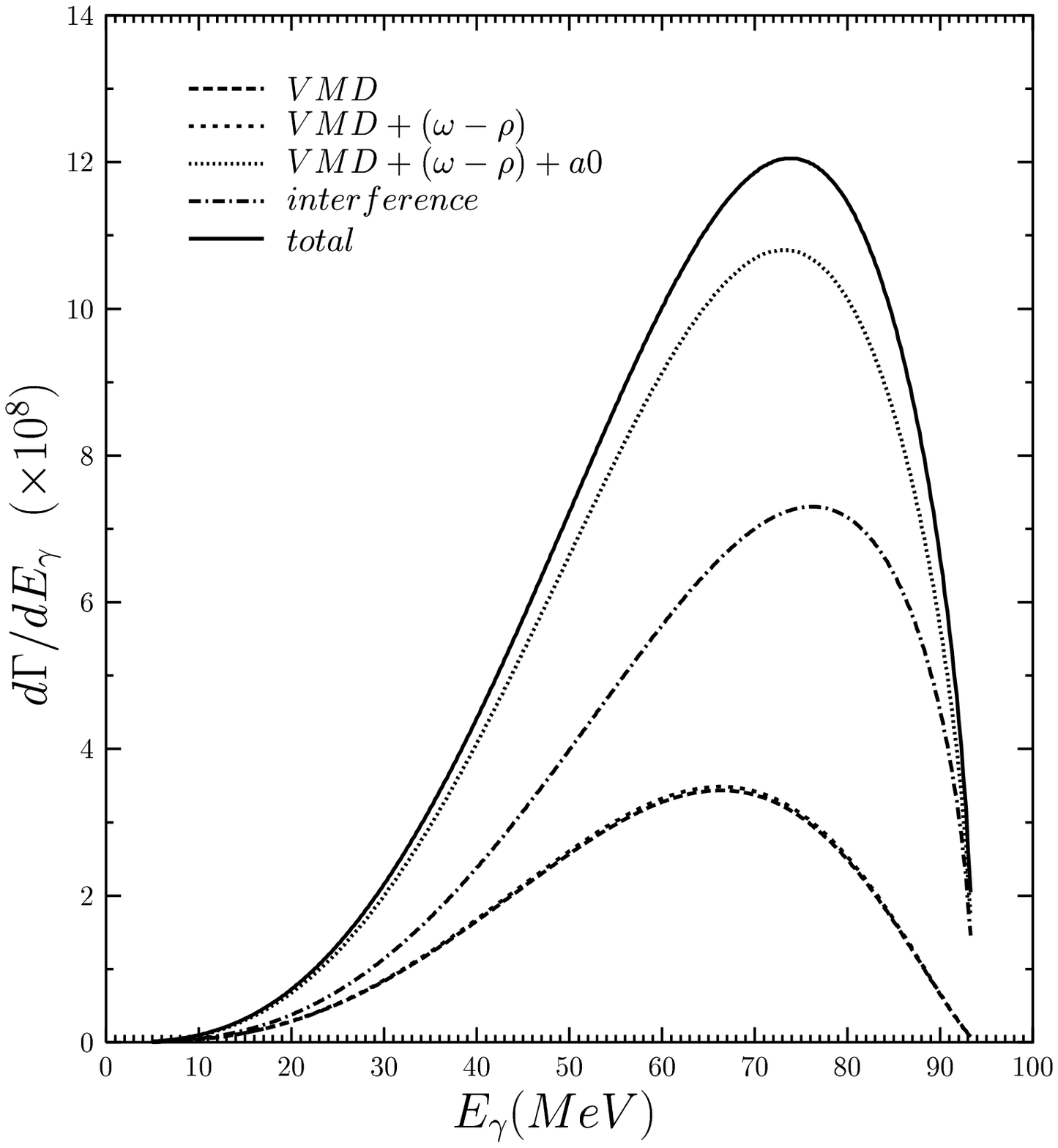,width=15cm,height=20cm}\vspace*{-3.0cm}
\caption{The photon spectra for the decay width of
$\omega\rightarrow\pi^{0}\eta\gamma$ decay. The contributions of
different terms resulting from the amplitudes of VMD, chiral
loops, $a_{0}$-meson intermediate state, and $\rho-\omega$ mixing
are indicated. } \label{fig3}
\end{figure}

\begin{figure}
\epsfig{figure=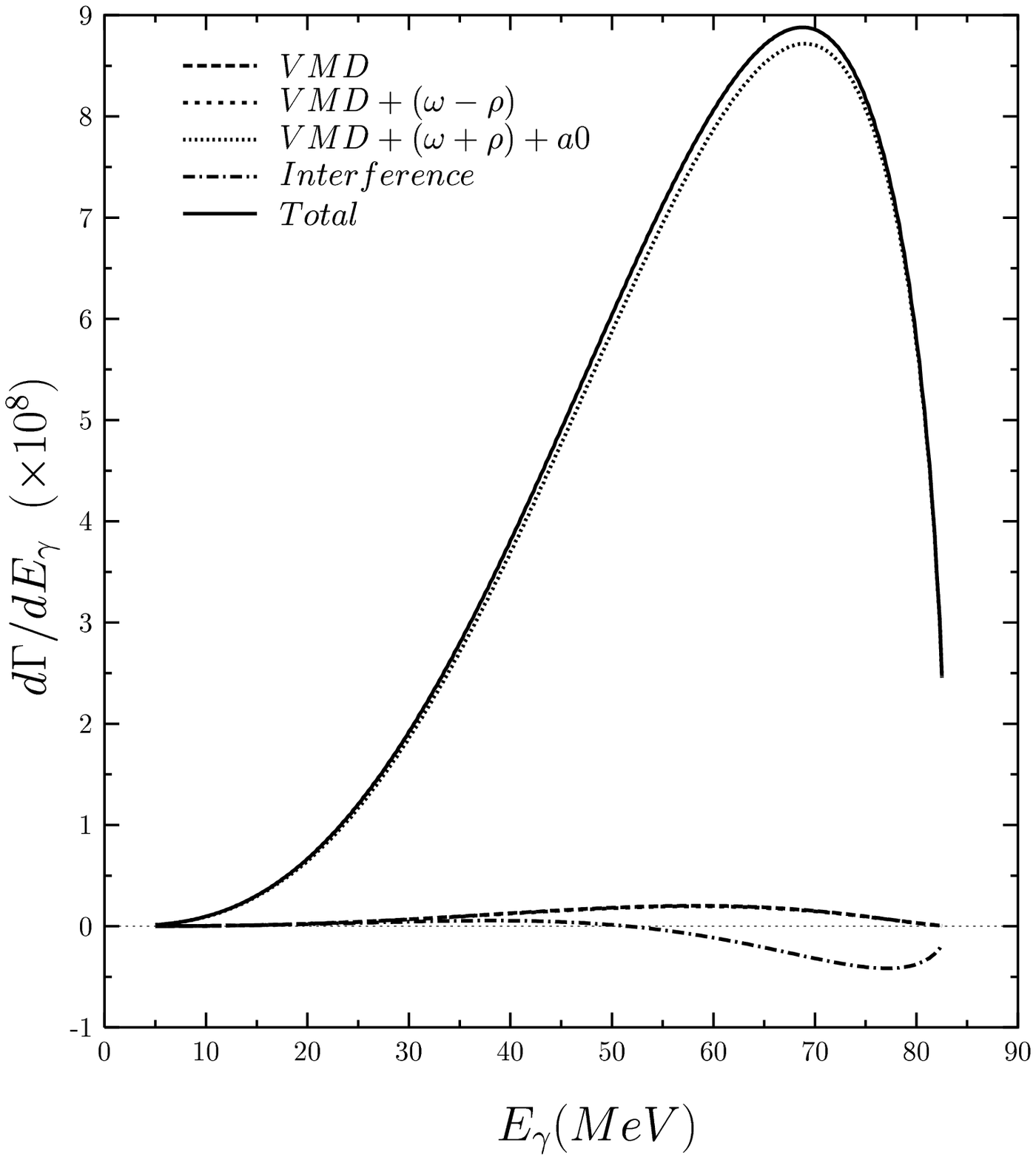,width=15cm,height=20cm} \vspace*{-3.0cm}
\caption{The photon spectra for the decay width of
$\rho\rightarrow\pi^{0}\eta\gamma$ decay. The contributions of
different terms resulting from the amplitudes of VMD, chiral
loops, $a_{0}$-meson intermediate state, and $\rho-\omega$ mixing
are indicated. } \label{fig4}
\end{figure}

\end{document}